\documentclass[12pt,a4paper]{article}
\usepackage{amsmath,amssymb,bm,geometry}
\title{Quantum, Stochastic, and Classical Dynamics Within A Single Geometric Framework}
\author{Partha Ghose\footnote{partha.ghose@gmail.com}\\{\small Tagore Centre for Natural Sciences and Philosophy, Kolkata, India}}
\date{}

\begin{document}
\maketitle

\begin{abstract}
Nelson's stochastic mechanics links quantum mechanics to an underlying Brownian motion with the identification $\hbar = m\sigma$. 
Ghose's interpolating equation introduces a continuous parameter $\lambda$ that suppresses the quantum potential $Q[\psi]$ and yields a smooth transition between quantum ($\lambda=0$) and classical ($\lambda=1$) regimes.
In this short note, we show that the Koopman--von Neumann (KvN) Hilbert-space formulation of classical mechanics emerges naturally as the $\lambda \to 1$ limit of this stochastic $\sigma$--$\lambda$ hierarchy.
The KvN phase-space amplitude provides an operator representation of the classical Liouville equation, while the $\lambda$ parameter acts as a projection flow from the complex projective Hilbert manifold $\mathbb{C}P^n$ to its classical quotient $\mathbb{C}P^*/U(1)$, implementing phase superselection. 
This unified picture links quantum, stochastic, and classical dynamics within a single continuous framework.
\end{abstract}

\section{Background}
In Nelson's stochastic mechanics \cite{Nelson1966}, the Schr\"{o}dinger equation emerges from forward--backward diffusion processes with diffusion coefficient $\sigma$ with the identification $\hbar=m\sigma$:

\begin{equation}
i m\sigma \frac{\partial \psi}{\partial t} = \left[-\frac{m\sigma^2}{2}\nabla^2 + V\right]\psi
\end{equation}
Guerra and Morato showed that the same equation can be derived using stochastic control theory \cite{Guerra-Morato} and that the classical Hamilton-Jacobi equation is modified by an additional stochastic term:  

\begin{equation}
\frac{\partial S}{\partial t} + \frac{(\nabla S)^2}{2m} + V + Q = 0 \label{hj}
\end{equation}
with $Q$ given by
\begin{equation}
Q = -\frac{m\sigma^2}{2}\frac{\nabla^2 \sqrt{\rho}}{\sqrt{\rho}},\,\,\,\,\psi = \sqrt{\rho}e^{iS/\hbar}
\end{equation}
In addition, one gets the continuity equation
\begin{equation}
\partial_t \rho + \nabla\!\cdot\!\!\left(\rho\,\frac{\nabla S}{m}\right) = 0.
\end{equation}

The Guerra-Morato variational approach works for any diffusion constant, and this can be taken care of by introducing a variable $\lambda$ ($0 \leq\lambda \leq 1$ ) and rewriting equation (\ref{hj}) in the form
\begin{equation}
\frac{\partial S}{\partial t} + \frac{(\nabla S)^2}{2m} + V + (1 -\lambda) Q = 0 \label{hjm}
\end{equation}
This corresponds to a modified Schrodinger equation
\begin{equation}
i\hbar \frac{\partial \psi}{\partial t} = \left[-\frac{\hbar^2}{2m}\nabla^2 + V - \lambda Q\right]\psi \label{eq:interpolating}
\end{equation}
which was first proposed  by Ghose (2002). The regime $\lambda=0$ corresponds to standard quantum mechanics, while $\lambda=1$ yields the ``classical Schr\"odinger'' or Rosen equation (Rosen 1964) whose polar decomposition reproduces the classical Hamilton--Jacobi and continuity equations.

\section{KvN Dynamics and Classical Limit}
Koopman \cite{Koopman} and von Neumann \cite{von Neumann} (KvN) showed that classical mechanics admits a Hilbert-space representation in which a complex amplitude $\Psi_{\mathrm{KvN}}(x,p,t)$ evolves under the Liouville operator:
\begin{equation}
i\,\partial_t \Psi_{\mathrm{KvN}}
 = \widehat{L}\Psi_{\mathrm{KvN}},\qquad
 \widehat{L} = \frac{\partial H}{\partial x}\frac{\partial}{\partial p}
              - \frac{\partial H}{\partial p}\frac{\partial}{\partial x}.
\label{eq:KvN}
\end{equation}
Writing $\Psi_{\mathrm{KvN}}=\sqrt{f}\,e^{i\phi}$ yields Liouville's equation $\partial_t f+\{H,f\}_{\mathrm{PB}}=0$
and a convective phase law $\partial_t\phi+\{H,\phi\}_{\mathrm{PB}}=0$. 
\emph{Phases are unobservable since classical observables act as multiplication by functions of $(x,p)$}; the theory is therefore subject to a phase superselection rule \cite{Sudarshan1976}.

Restricting to a Lagrangian sheet $p=\nabla S(x,t)$ with density $\rho(x,t)$ via
\begin{equation}
f(x,p,t)=\rho(x,t)\,\delta(p-\nabla S(x,t)),
\end{equation}
the KvN dynamics exactly reproduces the $\lambda=1$ polar equations
\begin{align}
\partial_t S + \frac{(\nabla S)^2}{2m} + V &= 0,
&
\partial_t \rho + \nabla\!\cdot\!\!\left(\rho\,\frac{\nabla S}{m}\right) &= 0.
\end{align}
Thus, the $\lambda=1$ limit of Eq.~\eqref{eq:interpolating} corresponds to KvN mechanics restricted to a single Lagrangian manifold.

\section{Embedding KvN in the $\sigma$--$\lambda$ Hierarchy}
The unified mapping can be summarized as
\begin{align}
&\text{Quantum regime ($\lambda=0$):}\quad &&\psi(x,t)=R e^{iS/\hbar},\ \ \hbar=m\sigma,\\
&\text{Interpolating ($0<\lambda<1$):}\quad &&i\hbar\,\partial_t\psi
 = \left[-\frac{\hbar^2}{2m}\nabla^2+V -\lambda Q[\psi]\right]\psi,\\
&\text{Classical limit ($\lambda=1$):}\quad &&\Psi_{\mathrm{KvN}}(x,p,t)\ \leftrightarrow\
 \psi(x,t)=\sqrt{\rho}\,e^{iS/\hbar},\ \ p=\nabla S.
\end{align}
In this picture, $\lambda$ acts as a continuous \emph{projection flow} from configuration-space Hilbert dynamics (quantum) to phase-space Hilbert dynamics (classical).
Define a phase-space density functional
\begin{equation}
F_{\sigma,\lambda}(x,p,t) = |\psi|^2\,\delta(p-\nabla S)
 + (1-\lambda)\,\eta_{\sigma}(x,p,t),
\end{equation}
where $\eta_{\sigma}$ encodes the diffusion-induced spread associated with $\sigma$. 
For $\lambda\to 1$, $\eta_{\sigma}$ vanishes, and $F_{\sigma,\lambda}$ reduces to the KvN amplitude projected on the Lagrangian sheet.

\subsection*{Geometric Interpretation}
The hierarchy can be viewed as a contraction of the quantum symplectic form
\begin{equation}
\omega_Q = \mathrm{d}x\wedge\mathrm{d}p + (1-\lambda)\,\omega_{\mathrm{quant}},
\end{equation}
where $\omega_{\mathrm{quant}}$ arises from the curvature associated with the quantum potential. 
As $\lambda\to 1$, the quantum term vanishes and the complex projective manifold $\mathbb{C}P^n$ contracts to the real symplectic phase space of KvN mechanics.
Identifying all states that differ by relative $U(1)$ phases---the quotient $\mathbb{C}P^*/U(1)$---implements Sudarshan's phase superselection and removes any residual interference.

\subsection*{Hierarchy Summary}
\begin{center}
\begin{tabular}{lll}
Regime & Character & Description \\\\ \hline
$\lambda=0$ & Fully quantum & Stochastic Schr\"odinger (Nelson) with $\hbar=m\sigma$ \\\\
$0<\lambda<1$ & Interpolating & Continuous suppression of $Q[\psi]$ (Ghose) \\\\
$\lambda=1$ & Classical & KvN Hilbert dynamics / Liouville flow with phase superselection \\\\
\end{tabular}
\end{center}
\section{Crossing, mixtures, and the CP$^*$ quotient}
In the $\lambda = 1$ limit, the wavefunction $\psi(x,t)=\sqrt{\rho}\,e^{iS/\hbar}$ defines a
single Lagrangian sheet $p=\nabla S$ in phase space. 
Trajectories generated by the velocity field $\mathbf{v}=\nabla S/m$ therefore never intersect;
this is the origin of the well-known ``no-crossing'' property of Bohmian paths.
Navia and Sanz~\cite{NaviaSanz2024} recently emphasized that, for the so-called
``classical Schr\"odinger equation,'' such non-crossing trajectories persist,
leaving a residual imprint of coherence that is foreign to true classical mechanics,
where different phase-space streams routinely intersect.

A genuinely classical ensemble should not be confined to a single phase sheet.
Classical multi-streaming is restored by admitting \emph{mixtures} of sheets,
\begin{equation}
f(x,p,t)
   = \sum_j w_j\,\rho_j(x,t)\,
     \delta\!\big(p-\nabla S_j(x,t)\big),
\label{eq:multisheet}
\end{equation}
where each component $(\rho_j,S_j)$ evolves according to the Hamilton--Jacobi and
continuity equations, and the weights $w_j$ form an incoherent statistical mixture.
In configuration space this corresponds to an ensemble of waves with distinct
phases $S_j$, permitting trajectory crossing at the ensemble level and eliminating
interference between components.

An equivalent and more compact formulation employs \emph{phase superselection}.
All states that differ only by local $U(1)$ phase factors are identified,
yielding the quotient manifold
\begin{equation}
CP^{*}=CP/U(1),
\end{equation}
as advocated by Sudarshan~\cite{Sudarshan1976}, who interpreted the phase as a hidden variable.
This identification enforces a superselection rule that removes any residual
interference among different $S_j$ branches, ensuring that observables depend
solely on the classical densities $\rho_j$ and momenta $\nabla S_j$.
In this quotient space the remaining dynamical variables coincide with those
of Koopman--von Neumann mechanics, and the ensemble~\eqref{eq:multisheet}
reproduces the correct classical Liouville behavior without any vestige of
quantum coherence.

\section{Concluding Remarks}
The $\sigma$--$\lambda$ hierarchy provides a natural bridge from quantum to classical mechanics in which KvN dynamics appears as the terminal point of a continuous stochastic deformation. 
It unifies stochastic mechanics, nonlinear interpolation, and Hilbert-space classical mechanics within a single geometric framework, offering a platform to explore controlled classicalization and phase-superselection phenomena.

\section{Acknowledgements}
This work resulted from a series of discussions with Anirudha Chakraborty, IIT Mandi. I am grateful to the Director, IIT Mandi for kind hospitality.


\begin{thebibliography}{9}
\bibitem{Nelson1966} E. Nelson, ``Derivation of the Schr\"{o}dinger Equation from Newtonian Mechanics,'' \emph{Phys. Rev.} \textbf{150}, 1079 (1966).
\bibitem{Guerra-Morato}
F. Guerra and L. M. Morato, ``Quantization of dynamical systems and stochastic control theory,'' \emph{Phys. Rev. D} \textbf{27}, 1774-1785 (1983).
\bibitem{Ghose2002} P. Ghose, ``A Continuous Transition Between Quantum and Classical Mechanics I,'' \emph{Found. Phys.} \textbf{32}, 871--906 (2002).
\bibitem{Rosen1964} N. Rosen, ``A Classical Analog of Quantum Mechanics,'' \emph{ Am. J. Phys.} \textbf{32}, 597 (1964).
\bibitem{Koopman}
B. O. Koopman, ``Hamiltonian Systems and Transformation in Hilbert Space,'' \emph{Proc. Natl. Acad. Sci. U.S.A.} \textbf{17}, 315-318 (1931).
\bibitem{von Neumann}
J. von Neumann, ``Zur Operatorenmethode in der Klassischen Mechanik,'' \emph{Ann. Math.} \textbf{33}, 587-642 (1932); ibid. \textbf{33}, 789-791 (1932).
\bibitem{Sudarshan1976} E. C. G. Sudarshan, ``Classical and Quantum Mechanics: A Unified Picture,'' \emph{Pramana} \textbf{6}, 117--126 (1976).

\bibitem{NaviaSanz2024} D. Navia and A. S. Sanz, ``Exploring the nonclassical dynamics of the
``classical'' Schr\"{o}dinger equation,'' \emph{Ann. Phys.} \textbf{463}, 169637 (2024); arXiv:2312.02977 (2024).
\end{thebibliography}
\end{document}